\documentclass[conference]{IEEEtran}%

\usepackage{amsmath,amsfonts,amssymb}
\usepackage{array}
\usepackage{bm}
\usepackage{subfigure}
\usepackage{textcomp}
\usepackage{stfloats}
\usepackage{url}
\usepackage{verbatim}
\usepackage{graphicx}
\usepackage{cite}
\usepackage{color}
\usepackage{optidef}
\usepackage{physics}
\usepackage[caption=false,font=normalsize,labelfont=sf,textfont=sf]{subfig}
\usepackage{algorithm}
\usepackage{algpseudocode}
\usepackage{etoolbox}
\usepackage{mathtools}

\newcommand\blfootnote[1]{%
  \begingroup
  \renewcommand\thefootnote{}\footnote{#1}%
  \addtocounter{footnote}{-1}%
  \endgroup
}

\begin{document}

\title{Denoising-Autoencoder-Assisted\\ Physical Layer Secret Key Generation}

\author{
\IEEEauthorblockN{Zhengyu Xu$^1$, Shun Kojima$^{1,2}$, and Shinya~Sugiura$^{1,*}$}
\IEEEauthorblockA{$^1$Institute of Industrial Science, The University of Tokyo\\
									$^2$Graduate School of Engineering Science, Yokohama National University\\
E-mail: \{jingyue8999@g.ecc.u-tokyo.ac.jp, skojima@iis.u-tokyo.ac.jp, sugiura@iis.u-tokyo.ac.jp\} 
}
}

\markboth{}
{Shell \MakeLowercase{\textit{et al.}}: A Sample Article Using IEEEtran.cls for IEEE Journals}

\IEEEpubid{}

\maketitle

\begin{abstract}
In this paper, we propose denoising autoencoder (DAE)-assisted secret key generation (SKG), where channel noise reciprocity imperfections induced due to wireless channel measurements are suppressed, hence significantly enhancing the reliability and efficiency. More specifically, the DAE is capable of capturing the intrinsic structure of input distributions, reconstructing the original data structure, and removing additive noise while preserving the essential structure of signals. In our analysis, it is demonstrated that the proposed SKG scheme exhibits higher performance than the conventional schemes in terms of key disagreement rate (KDR), secret key capacity (SKC), and randomness of the generated keys.
\end{abstract}

\section{Introduction}
\label{sec:Introduction}
\blfootnote{Preprint for publication in \textit{IEEE International Conference on Communications (ICC)}, Glasgow, Scotland, UK, May 2026, DOI: 10.1109/ICC59461.2026.11587401. $\copyright$ 2026 IEEE. Personal use of this material is permitted. Permission from IEEE must be obtained for all other uses, in any current or future media, including reprinting/republishing this material for advertising or promotional purposes, creating new collective works, for resale or redistribution to servers or lists, or reuse of any copyrighted component of this work in other works.}

\IEEEPARstart{T}{he} Internet of Things (IoT) has witnessed rapid development and applications, including smart homes, healthcare, transportation, industrial automation, and disaster response, and significantly enhanced the quality of life. 
However, the broadcast nature of wireless channels is inherently vulnerable to security threats of eavesdropping and jamming~\cite{Zou2016}. 
Recent development of quantum computing technology poses a significant threat to widely employed public-key cryptosystems, including Rivest-Shamir-Adleman (RSA) and elliptic curve cryptography~\cite{Mukherjee2014}. Hence, in order to maintain quantum-resistant high security in the secret key agreement and authentication based on the current public-key cryptosystems, substantial complexity and overhead are imposed~\cite{10872957}. Furthermore, due to the growing cyber attacks, conventional public key infrastructure-based schemes may not fulfill the security requirements in rapidly expanding distributed and dynamic networks~\cite{7547899}.

To combat the above-mentioned limitations imposed on the confidential secret key agreement, secret key generation (SKG) has emerged as a promising alternative to conventional public-key cryptosystems~\cite{zhang2016key}. In the SKG system, physical randomness inherent in wireless channels is exploited to generate shared secret keys, which may be used for information-theoretic secure communication with the aid of a one-time pad, unlike traditional cryptographic systems that rely on computational security. More specifically, the reciprocity and temporal variation of the wireless channel between two legitimate nodes are used to extract common randomness. Typically, radio waves experience scattering, reflection, and diffraction~\cite{goldsmith}, which decorrelates the eavesdropper's channel from the legitimate channel when the eavesdropper is positioned sufficiently apart from the legitimate users. This spatial decorrelation allows us to exploit physical-layer SKG without relying on the unreliable assumption of computational security and to offer inherent resilience against quantum attacks~\cite{zhang2016efficient}.

However, in practical wireless environments, the assumption of perfect channel reciprocity between the legitimate users may not be satisfied due to synchronization errors, environmental fluctuations, external interference, and hardware impairments.
Then, an unignorable mismatch between the channel coefficients estimated at the legitimate users induces the increased disagreement of the generated secret key sequences~\cite{Liu2012RSSKey}. There have been diverse SKG schemes that aim for improving the achievable performance~\cite{related1,related5,related7,7997419,Lin2020Efficient,10440494,related13,related15,11086605}.
While conventional statistical signal processing and filtering schemes have limited benefits from addressing such issues,  sophisticated filtering schemes, pilot designs, and preprocessing have been developed for improving the SKG performance~\cite{related1,related5,related7,7997419,Lin2020Efficient}.
More recently, there have been developed several machine learning (ML)-based SKG schemes~\cite{10440494,related13,related15,11086605},
which exhibit promise for correcting channel measurements obtained by the legitimate users.

Against the above-mentioned background, the novel contribution of this paper is to propose a denoising autoencoder (DAE)-assisted SKG for the first time, where a neural network reconstructs the channels in a manner to be highly correlated among the legitimate users.
More specifically, the proposed scheme captures intrinsic structures of input data through artificial noise injection and subsequent reconstruction. As a result, reciprocity imperfections are mitigated and the additive noises are suppressed, hence improving the achievable key agreement performance. Furthermore, we provide our performance results to demonstrate that the proposed scheme is capable of achieving higher performance than the conventional benchmarks, i.e., discrete wavelet transform (DWT)~\cite{7847039}, principal component analysis (PCA)~\cite{related7}, discrete cosine transform (DCT)~\cite{7997419}, and traditional autoencoder (AE)~\cite{related15}, in terms of key disagreement rate (KDR), secret key capacity (SKC), and the statistical randomness of the generated keys.

\section{System Model}
\label{sec:SystemModel}
Let us consider the scenario with the two legitimate users, Alice and Bob, who aim to share a secret key based on SKG.
We assume time-division duplex to exchange pilot symbols, where the same frequency band is used to allow a high correlation between the link from Alice to Bob and that from Bob to Alice. We also consider the presence of a single passive eavesdropper, Eve, who tries to intercept the pilot exchange between Alice and Bob.
For the sake of simplicity but without loss of generality, we assume that Eve is located close to Alice and sufficiently apart from Bob according to~\cite{10379509, related5, 9217320, 7809064}, where Eve intercepts the generated secret key from the pilot symbols transmitted from Alice.

We assume frequency-flat Rayleigh fading for the channel for each link.
Here, the channel impulse response $\xi(t)$ is expressed as
$\xi(t) = \sum_{l=1}^{L} a_l\, \delta(t - \tau_l)$,
where $L$ denotes the number of multi-paths, while $a_{l}$ and $\tau_{l}$ represent the complex-valued channel coefficient and propagation delay of the $l$-th path, respectively. Furthermore, $\delta(t)$ is the Dirac delta function.

The legitimate users transmit orthogonal frequency-division multiplexing (OFDM)-based pilot symbols, having the $N_c$ subcarriers with the center frequencies of $f_k = f_0 + k \Delta f$ for $k = 0,\dots, N_c-1$, where $\Delta f$ denotes the subcarrier spacing.
The frequency response vector $\mathbf{h}\in\mathbb{C}^{N_c}$ across all the subcarriers is represented by
$\mathbf{h}
= \left[h^{(0)}, \cdots, h^{(N_c-1)}\right]^T$,
where
$h^{(k)}
= \int_{-\infty}^{\infty} \xi(\tau) \, e^{-j 2\pi f_k \tau} \, d\tau$.

Moreover, we consider the imperfect channel reciprocity model as follows~\cite{6198402}:
\begin{IEEEeqnarray}{rCL}
h_{AB}^{(k,l)} &=& \rho_{AB}h_{BA}^{(k,l)} + \sqrt{1-\rho_{AB}^2}w_{AB}^{(k,l)} \label{eq:correlation}\\
h_{BE}^{(k,l)} &=& \rho_{BE}h_{BA}^{(k,l)} + \sqrt{1-\rho_{BE}^2}w_{BE}^{(k,l)}.
\end{IEEEeqnarray}
where $k \in \{0,1,\dots, N_c-1\}$ denotes the subcarrier index, and $l \in \{0,1,\dots, N_f-1\}$ is the frame index.
Furthermore, $h_{BA}^{(k,l)}$, $h_{AB}^{(k,l)}$, and $h_{BE}^{(k,l)}$ represent the channel frequency response from Bob to Alice, that from Alice to Bob, and that from Bob to Eve, respectively. Also, $w_{AB}^{(k,l)}$ and $w_{BE}^{(k,l)}$ are the associated independent and identically distributed circularly symmetric complex-valued Gaussian random variables with zero mean and the same variance as that of $h_{BA}^{(k,l)}$.
Furthermore, $\rho_{AB}$ and $\rho_{BE}$ are the correlations of bidirectional channels between Alice and Bob, as well as that between Bob and Eve, respectively, which are defined as follows:
\begin{IEEEeqnarray}{rCL}
&\rho_{XY} = {
\mathbb{E}\left[\,h_{XY}^{(k,l)}\,h_{YX}^{(k,l)*}\right]
}/
{
\sqrt{
\mathbb{E}\left[\left|h_{XY}^{(k,l)}\right|^2\right] \,
\mathbb{E}\left[\left|h_{YX}^{(k,l)}\right|^2\right]
}
} & \label{eq:correlation_def} \\
&\quad X, Y \in \{A, B, E\}, \ \  k = 0, \ldots, N_c - 1, \ \ l=0,\cdots,N_f.& \nonumber
\end{IEEEeqnarray}
where ${}^*$ indicates the complex conjugate, and $\mathbb{E}[\cdot]$ represents the expectation operator.

\section{Proposed DAE-Assisted SKG}
\label{sec:Proposedscheme}
This section proposes the novel DAE-assisted SKG scheme, which improves key agreement between the legitimate users despite the presence of the imperfect channel reciprocity and additive noises.

\begin{figure}[t]
\centering
\includegraphics[width=.7\columnwidth]{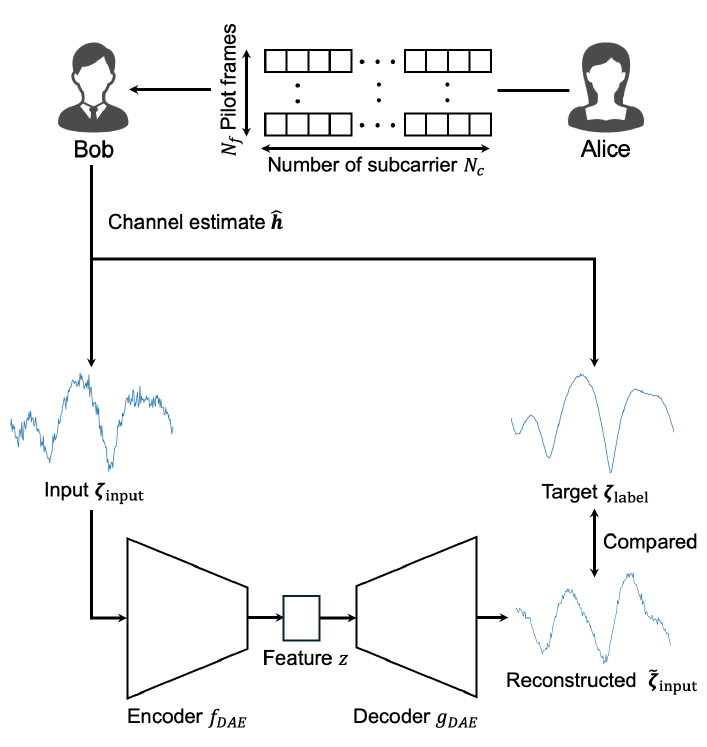}
\caption{The offline training procedure for the DAE.}
\label{fig:DAEtraining}
\end{figure}

The proposed DAE-assisted SKG scheme is composed of an offline training phase (Fig.~\ref{fig:DAEtraining}) and an online SKG phase.
More specifically, in the offline training phase, channel probing samples are collected to optimize the DAE parameters.
Once the offline training phase is completed, the optimized DAE parameters are shared among the legitimate users.
Then, the legitimate users share a secret key sequence by implementing the online SKG phase, including channel probing,  machine learning-based filtering,  quantization, and information reconciliation.

\subsection{Offline Training Phase}
\label{subsec:Offline Training Phase}

\subsubsection{Preparation of Training Dataset}
\label{subsubsec:Preparing DAE Training Dataset}

First, the dataset is generated according to our numerical multi-path fading models. Note that successful convergence of the DAE requires noiseless reference signals, while sufficient randomness as the entropy source of SKG has to be achieved through multi-path channels.
$N_{DAE}$ estimated channel sample vector pairs $\hat{\mathbf{h}}_\mathrm{input}\in\mathbb{C}^{N_c}$ and $\hat{\mathbf{h}}_\mathrm{target}\in\mathbb{C}^{N_c}$ are generated according to a channel estimation model considered. More specifically, $\hat{\mathbf{h}}_\mathrm{input}$ and $\hat{\mathbf{h}}_\mathrm{target}$ are contaminated by the independent AWGNs with the SNRs of $\gamma_\mathrm{input}$ and $\gamma_\mathrm{target}$, respectively, while having the same channels before estimation.
Then, $\hat{\mathbf{h}}_\mathrm{input}$ and $\hat{\mathbf{h}}_\mathrm{target}$ are converted to RSS ones as $\boldsymbol{\zeta}_\mathrm{input}=|\hat{\mathbf{h}}_\mathrm{input}|$ and $\boldsymbol{\zeta}_\mathrm{target}=|\hat{\mathbf{h}}_\mathrm{target}|$.

\subsubsection{DAE Training}
\label{subsubsec:DAE Training}

In our DAE training, we employ an autoencoder consisting of an $D$-layer encoder $f_{\mathrm{DAE}}$ and a symmetric $D$-layer decoder $g_{\mathrm{DAE}}$. The encoder maps the input RSS dataset $\boldsymbol{\zeta}_{\mathrm{input}}$ to a low-dimensional latent feature vector $\mathbf{z}\in\mathbb{R}^{N_c/2^D}$. The entire encoding process is denoted by $\mathbf{z} = f_{\mathrm{DAE}}(\boldsymbol{\zeta})$. More specifically, each encoder layer consists of two functions, i.e., linear transformation and nonlinear activation, designed for performing dimensional reduction.

The $d$-th layer of the encoder operates as follows:
\begin{IEEEeqnarray}{rCL}
\mathbf{y}^{(d)} = s^{(d)}\bigl(\mathbf{W}^{(d)} \mathbf{y}^{(d-1)} + \mathbf{b}^{(d)}\bigr) \in \mathbb{R}^{N_c/2^d} \ \textrm{for} \ d=1,\cdots,D, \nonumber
\end{IEEEeqnarray}
where $\mathbf{y}^{(d)}$ represents the output of the $d$-th encoder layer with the initial values of $\mathbf{y}^{(0)} = \boldsymbol{\zeta}$.
Furthermore, $s^{(d)}(\cdot)$, $\mathbf{W}^{(d)}\in\mathbb{R}^{N_c/{2^{d}} \times N_c/{2^{(d-1)}}}$, and $\mathbf{b}^{(d)}\in\mathbb{R}^{N_c/2^d}$ denote the activation function, the weight matrix, and the bias vector at the $l$-th layer, respectively.
The output of the $D$-th layer is the latent feature vector $\mathbf{y}^{(D)} = \mathbf{z}$.

Next, the latent feature vector $\mathbf{z}$ is fed into the decoder $g_{\mathrm{DAE}}$ to obtain the reconstructed RSS output as $\tilde{\boldsymbol{\zeta}} = g_{DAE}(\mathbf{z}) \in\mathbb{R}^{N_c}$. The $l$-th layer of the decoder outputs:
\begin{IEEEeqnarray}{rCL}
\tilde{\mathbf{y}}^{(d)} = \tilde{s}^{(d)} \bigl(\tilde{\mathbf{W}}^{(d)} \tilde{\mathbf{y}}^{(d-1)} + \tilde{\mathbf{b}}^{(d)}\bigr)  \ \textrm{for} \ d=1,\cdots,D,
\label{eq:decoder}
\end{IEEEeqnarray}
where $\tilde{s}^{(d)}(\cdot)$, $\tilde{\mathbf{W}}^{(d)}\in\mathbb{R}^{N_c/{2^{(D-d)}} \times N_c/{2^{(D-d+1)}}}$, and $\tilde{\mathbf{b}}^{(d)}\in\mathbb{R}^{N_c/2^{D-d}}$ denote the activation function, the weight matrix, and the bias vector in the $d$-th decoder layer, respectively, while we have the initial input $\tilde{\mathbf{y}}^{(0)} = \mathbf{z}$ for the decoder. Hence, the output at the decoder is given by $\tilde{\boldsymbol{\zeta}} = \tilde{\mathbf{y}}^{(D)}$.

Finally, the loss function to be minimized for training the DAE's parameters of ($\mathbf{W}^{(d)}$, $\tilde{\mathbf{W}}^{(d)}$, $\mathbf{b}^{(d)}$, $\tilde{\mathbf{b}}^{(d)}$) is given by the MSE between the reconstructed output $\tilde{\boldsymbol{\zeta}}$ at the decoder and the target ${\boldsymbol{\zeta}}_\mathrm{target}$ as follows:
$\mathcal{L}
= \mathbb{E}\left[\left\| \boldsymbol{\zeta}_\mathrm{target} - \tilde{\boldsymbol{\zeta}} \right\|^{2}\right]$.
\subsection{Online SKG Phase}
\label{subsec:Online Secret Key Generation Phase}
\subsubsection{Channel Probing}
\label{subsubsec:Channel Probing}

In the SKG phase, the legitimate users exchange pilot symbols to acquire the channel coefficients.
The signal received by user Y from user X during the $l$-th pilot frame is distorted by the channel and corrupted by the AWGN.
Assuming a known frequency-domain pilot symbols $\mathbf{x} \in \mathbb{C}^{N_c}$, the received vector $\mathbf{y}_{XY}^{(l)}$ is given by~\cite{related5}:
\begin{IEEEeqnarray}{rCL}
&\mathbf{y}_{XY}^{(l)} = \mathbf{h}_{XY}^{(l)} \odot \mathbf{x} + \mathbf{n}_{XY}^{(l)} \in\mathbb C^{N_c}&  \\
&\quad X, Y \in \{A, B, E\}, \ \  l = 0, \ldots, N_f - 1,& \nonumber
\end{IEEEeqnarray}
where
$\mathbf{h}_{XY}^{(l)} = \left[h_{XY}^{(0,l)}, \dots, h_{XY}^{(N_c-1,l)}\right]^T$.
Also, $\mathbf n_{AB}^{(l)}$, $\mathbf n_{BA}^{(l)}$, and $\mathbf n_{BE}^{(l)}$ are the associated AWGN components, having the variance $\sigma_0^{2}$ and $\odot$ denotes the Hadamard product, i.e., the element-wise product.
The receiver performs least squares estimation~\cite{9298937} to achieve the estimated channels $\hat{\mathbf{h}}_{XY}^{(l)}$ from ${\mathbf{y}_{XY}^{(l)}}$ and ${\mathbf{x}}$. 
Then, the channel estimates are averaged over the entire $N_f$ pilot frames to obtain
$\bar{\mathbf h}_{XY} = \frac{1}{N_f}\sum_{l=0}^{N_f-1} \hat{\mathbf{h}}_{XY}^{(l)} \in \mathbb{C}^{N_c}$.

\subsubsection{ML-Based Filtering}
\label{subsubsec:ML-based Filtering}
The channel estimates $\hat{\mathbf{H}}_{XY}$  $(X, Y \in \{A, B, E\})$ includes the effects of are imperfect channel reciprocity. To suppress this degradation, we introduce an ML-based filtering stage based on offline offline-trained DAE.

More specifically, the RSS of $\bar{\mathbf h}_{XY}$ is input to our DAE as
\begin{IEEEeqnarray}{rCL}
\mathbf{z}_{XY}=f_{\mathrm{DAE}}\left(\bar{\boldsymbol{\zeta}}_{XY}\right), \label{eq:z-xy}
\end{IEEEeqnarray}
where
$\bar{\boldsymbol{\zeta}}_{XY} =\left|\bar{\mathbf h}_{XY}\right| \in \mathbb{R}^{N_c}$.
Then, $\mathbf{z}_{XY}$ of \eqref{eq:z-xy} is input to our DAE decoder
$g_{\mathrm{DAE}}$ to reconstruct the de-noised RSS of the channel estimates $\tilde{\boldsymbol{\zeta}}_{XY}$ from the latent representation as follows:
$\tilde{\boldsymbol{\zeta}}_{XY} = g_{\mathrm{DAE}}(\mathbf{z}_{XY})$.

\begin{figure}[!t]
\centering
\subfigure[Without DAE]{\includegraphics[width=55mm]{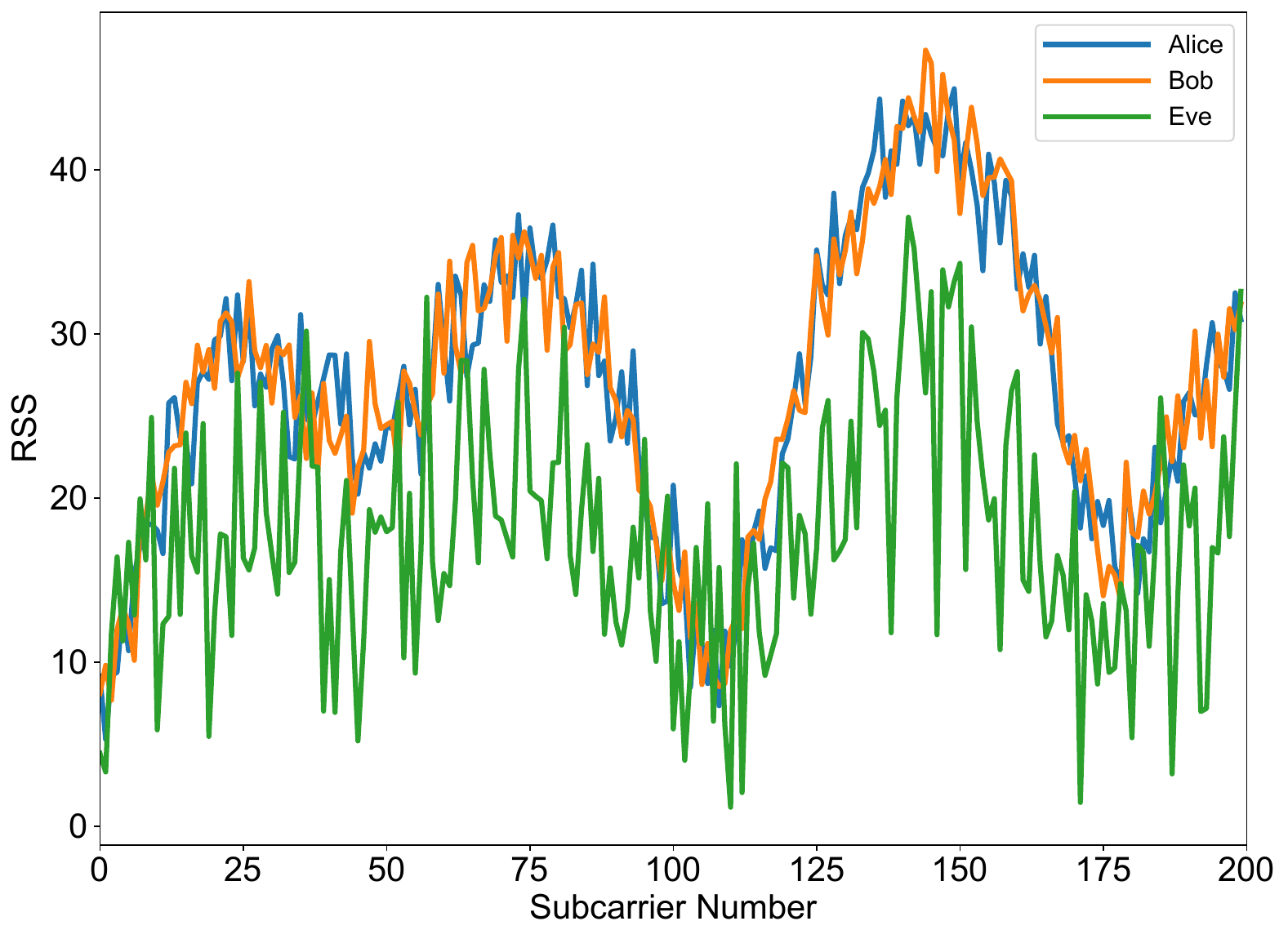}}\label{fig:fig4a}\\
\subfigure[With DAE]{\includegraphics[width=55mm]{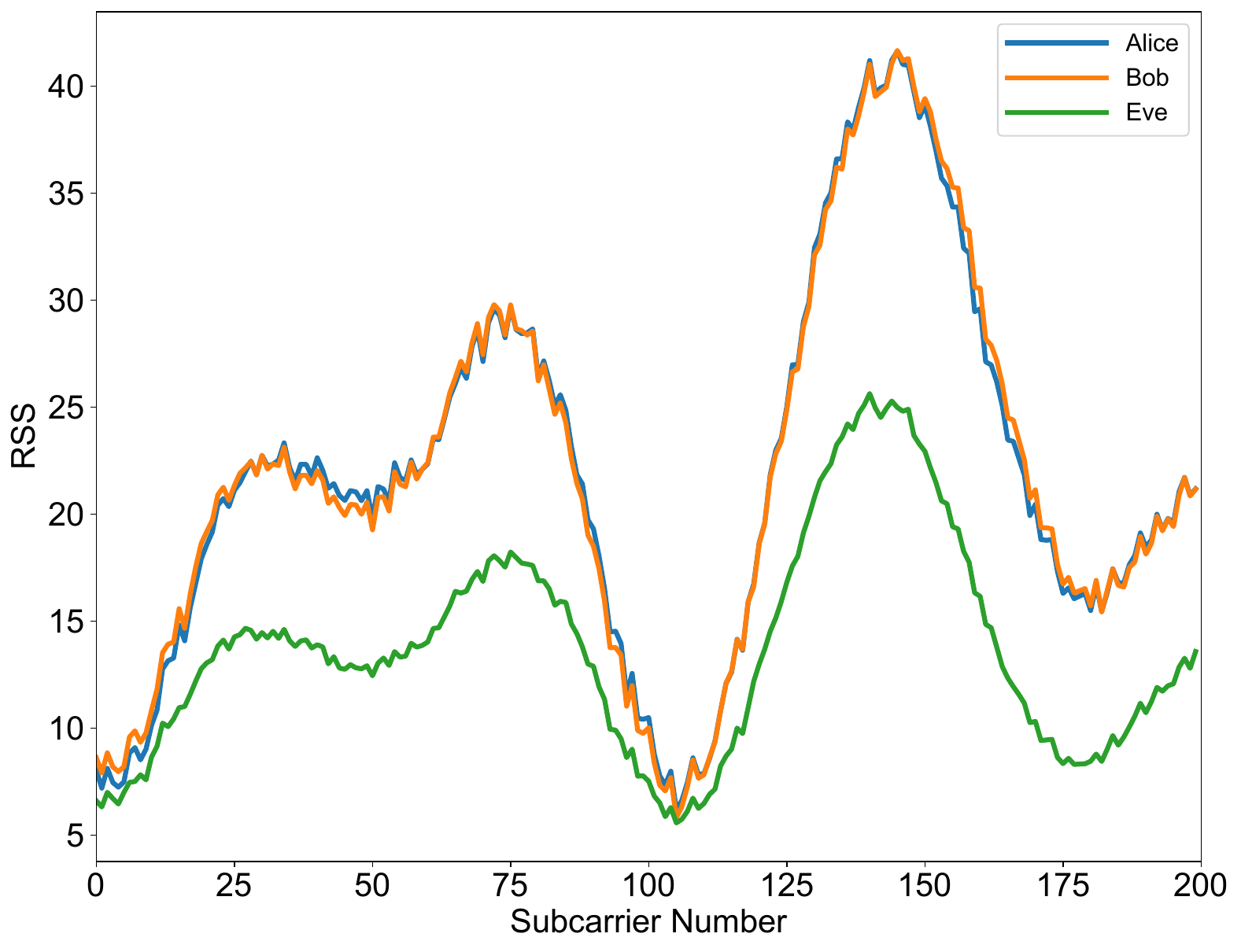}}\label{fig:fig4b}
\caption{The RSS of the estimated channels at Alice, Bob, and Eve with the parameters of $\rho_{BA}=1$, $\rho_{BE}=0.6$, and the SNR $=10$~dB.}
\label{fig:reconstruct_rss}
\end{figure}

In Fig.~\ref{fig:reconstruct_rss}, let us exemplify the results of DAE-based filtering, where we considered $\rho_{AB}=1$, $\rho_{BE}=0.6$, $N_f = 5$, and $N_c = 512$. In our training,  $N_{DAE}=10,000$ estimated channel sample vector pairs $\hat{\mathbf{h}}_\mathrm{input}$ and $\hat{\mathbf{h}}_\mathrm{target}$ are generated with the SNRs of $\gamma_\mathrm{input}=0$~dB and $\gamma_\mathrm{input}=30$~dB, respectively. Based on each pair, the DAE model is trained offline with $1,000$ epochs. Fig.~\ref{fig:fig4a} shows the estimated channels without DAE-filtering At Alice, Bob, and Eve with SNR $=10$~dB, while Fig.~\ref{fig:fig4b} exhibits the corresponding DAE-filtered reconstruction counterparts.
It is found that the RSSs of Alice and Bob exhibited an improved similarity, which reduces the effects of noise.

\subsubsection{Quantization}
\label{subsubsec: Quantization}

Once the reconstructed channel estimates are obtained in the form of the RSS, multilevel quantization~\cite{Premnath2013} is applied to convert the RSS values into a bit sequence.
First, the empirical distribution of the RSS is analyzed to determine its dynamic range and characteristics. First, the empirical distribution of the RSS is analyzed to determine its dynamic range and key characteristics.
Then, the RSS range is divided into $L$ intervals using $L-1$ thresholds, which are defined by the quantile function $Q(\cdot)$ of the cumulative distribution function as
$\lambda_i = Q\left({i}/{L}\right)$ $(i = 1, \dots, L-1)$,
where $\lambda_i$ denotes the $i$-th linear threshold. Each sample is then mapped to a $\log_2L$ bits.

\subsubsection{Information Reconciliation}
Finally, information reconciliation is carried out to improve the key agreement between the legitimate users. In this paper, we employ Bose–Chaudhuri–Hocquenghem (BCH) code-based information reconciliation~\cite{BCH}.
More specifically, Alice encodes a quantized bit sequence $\mathbf{k}_A \in \{0,1\}^k$ using the BCH encoder to obtain the codeword of
$\mathbf{m}_A = \left[\mathbf{k}_A^T \ \mathbf{p}^T\right]^T$,
where $\{\cdot \mid \cdot\}$ denotes row-wise concatenation, $\mathbf{p} \in \{0,1\}^{n-k}$ represents the parity bits, and $\mathbf{k}_A \in \{0,1\}^n$ is the encoded codeword. Then the parity component $\mathbf{p}$ is transmitted to Bob over a public channel.

Upon receiving the parity bits $\mathbf{p}$, Bob concatenates them with his own quantized bit sequence $\mathbf{k}_B \in \{0,1\}^k$ to form the candidate codeword as follows:
$\mathbf{m}_B = \left[\mathbf{k}_B^T \ \mathbf{p}^T\right]^T$,
which serves as the input to the BCH decoder. Bob then decodes $\mathbf{c}_B$ to recover Alice's key $\mathbf{k}_A$, hence correcting any bit errors between the two sequences.

\section{Performance Results}
\label{sec:SimulationResults}

\subsection{Simulation Setup}
\label{subsec:SimulationSetup}

In our simulations, random-phase 15 paths with exponential decay were generated to represent multi-path fading channels.
The maximum Doppler frequency was set to 1~Hz, and a single antenna element was employed at each node.
\begin{table}[t]
\footnotesize
\caption{DAE Architecture}
\label{tab:dae_model}
\begin{center}
\renewcommand{\arraystretch}{1.2}
\begin{tabular}{lccc} \hline
\textbf{Layer}    & \textbf{Input size} & \textbf{Output size} & \textbf{Activation function} \\ \hline
1st Encoder layer & 512 & 256                       & ReLU                          \\ 
2nd Encoder layer	& 256 & 128                       & ReLU                          \\ 
1st Decoder layer & 128 & 256                       & ReLU                          \\ 
2nd Decoder layer	& 256 & 512                       & Linear                        \\ \hline
\end{tabular}
\end{center}
\end{table}

The architecture of the proposed DAE is summarized in Table~\ref{tab:dae_model}.
The DAE consists of $D=2$ encoder layers and $D=2$ decoder layers, where the input data size is set to $N_c=512$.
Each of the two encoder layers compresses the data size from 512 to 256 and from 256 to 128, respectively. At each of the two decoder layers, the compressed data size is restored from 128 to 256 and finally from 256 to 512.
The rectified linear unit (ReLU) activation function is used for the two encoder layers and for the hidden decoder layer, while the linear activation is employed for the output decoder layer.

Furthermore, the DAE hyperparameters are listed in Table~\ref{tab:dae_params}.
To achieve fast and stable convergence, the Adam optimizer~\cite{2015-kingma} is employed for our DAE training, which adjusts the learning rate in an adaptive manner, based on the first- and second-order moment estimates of the gradients. The weights are initialized using Xavier uniform initialization~\cite{pmlr-v9-glorot10a}, and the initial bias terms are set to zeros in order to prevent unintended shifts of the model outputs.
Also, the exponential decay rates for the moment estimates in the Adam optimizer are set to $\beta_1 = 0.9$ and $\beta_2 = 0.999$. The training dataset comprises 80\% of the total samples, while the remaining 20\% are reserved for validation.
The model is trained for 1,000 epochs.

Unless otherwise noted, we employ the parameters of $N_f = 1$,  $N_c = 512$, and $L = 8$. In our offline DAE training, the input data is generated for SNR $=0$~dB, while the target data is generated for SNR $=30$~dB.

\begin{table}[t]
\footnotesize
\caption{DAE Parameters}
\label{tab:dae_params}
\begin{center}
\renewcommand{\arraystretch}{1.2}
\begin{tabular}{lc} \hline
\textbf{Parameter}                             & \textbf{Value}                          \\ \hline
Optimizer                                      & Adam \cite{2015-kingma}                 \\ 
Kernel initializer                             & glorot uniform \cite{pmlr-v9-glorot10a} \\ 
Bias initializer                               & zeros                                   \\ 
Exponential decay rate $\beta_1$~\cite{2015-kingma}             & 0.9                                     \\ 
Exponential decay rate $\beta_2$~\cite{2015-kingma}             & 0.999                                   \\ 
Number of epochs                               & 1,000                                   \\ 
Number of training samples                     & 8,000                                   \\ 
Number of validation samples                   & 2,000                                   \\ \hline
\end{tabular}
\end{center}
\end{table}

\subsection{SKG Benchmarks}
In our simulations, we consider the five benchmark schemes, i.e.,
the naive SKG scheme without preprocessing, denoted as the `direct' SKG scheme, as well as the transform-based SKG schemes aided by the DWT~\cite{Lin2020Efficient}, the DCT~\cite{7997419}, the PCA~\cite{related7}, and the AE~\cite{related15}. Each of the schemes based on DWT, the DCT, and the AE employs a specific transform of the RSS values of the estimated channels, while the PCA scheme implements the transform of the complex-valued estimated channels.

\subsection{KDR Performance}
\label{sec:EvaluationofKDR}
To evaluate the reliability of the proposed scheme, we compared the KDRs, which correspond to the error rate between the secret keys generated at Alice and Bob.
To be more specific, the KDR is represented by
$P_{AB} = \frac{1}{N_k} \sum_{j=1}^{N_k} \bigl|\mathbf{k}_{A}(j) - \mathbf{k}_{B}(j)\bigr|$,
where $\mathbf{k}_{A}(j)$ and $\mathbf{k}_{B}(j)$ are the $j$-th bits of the key sequences, generated at Alice and Bob, respectively, while $N_k$ is the total length.

\begin{figure}[!t]
\centering
\includegraphics[width=.8\columnwidth]{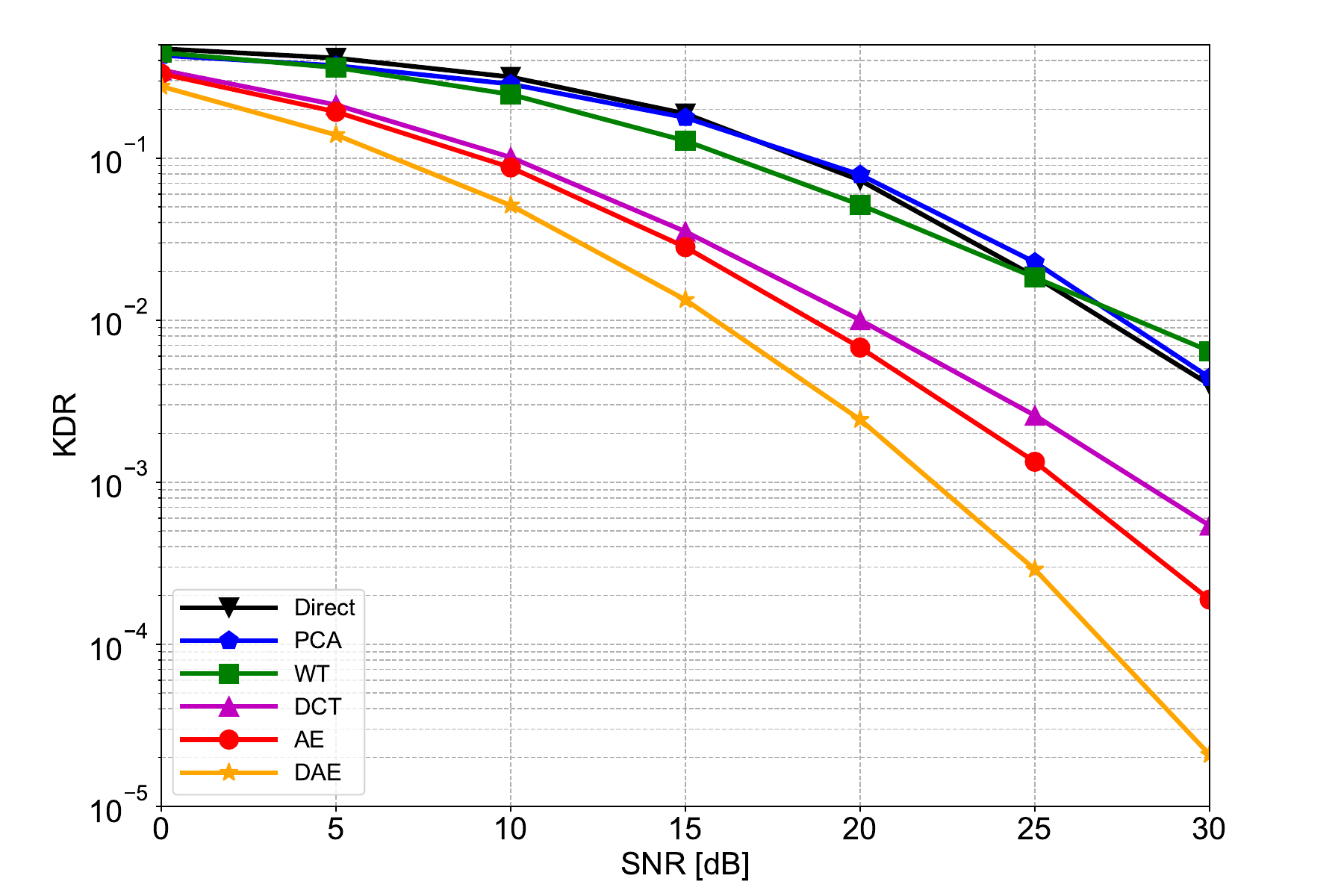}
\caption{Comparisons of the KDRs between the proposed scheme and the five benchmarks for $\rho_{AB} = 1$.}
\label{fig:kdr_fig1}
\end{figure}

Fig.~\ref{fig:kdr_fig1} compares the KDRs for the channel correlation of $\rho_{AB} = 1$, where, based on \eqref{eq:correlation}, perfect channel reciprocity is assumed between Alice and Bob except for the effects of AWGNs.
Observe in Fig.~\ref{fig:kdr_fig1} that upon increasing the SNR, the KDR of each scheme monotonically decreases, and the proposed DAE-assisted SKG scheme outperforms all the benchmarks.

\subsection{MI Performance}
\label{sub:Evaluation of MI}

We evaluate the MI $I(\tilde{\boldsymbol{\zeta}}_{BA}; \tilde{\boldsymbol{\zeta}}_{AB})$ of the proposed scheme between the reconstructed RSS values $\tilde{\boldsymbol{\zeta}}_{BA}$ and $\tilde{\boldsymbol{\zeta}}_{AB}$ based on the $k$-nearest neighbors algorithm~\cite{dasarathy1991nearest}.

\begin{figure}[!t]
\centering
\includegraphics[width=.8\columnwidth]{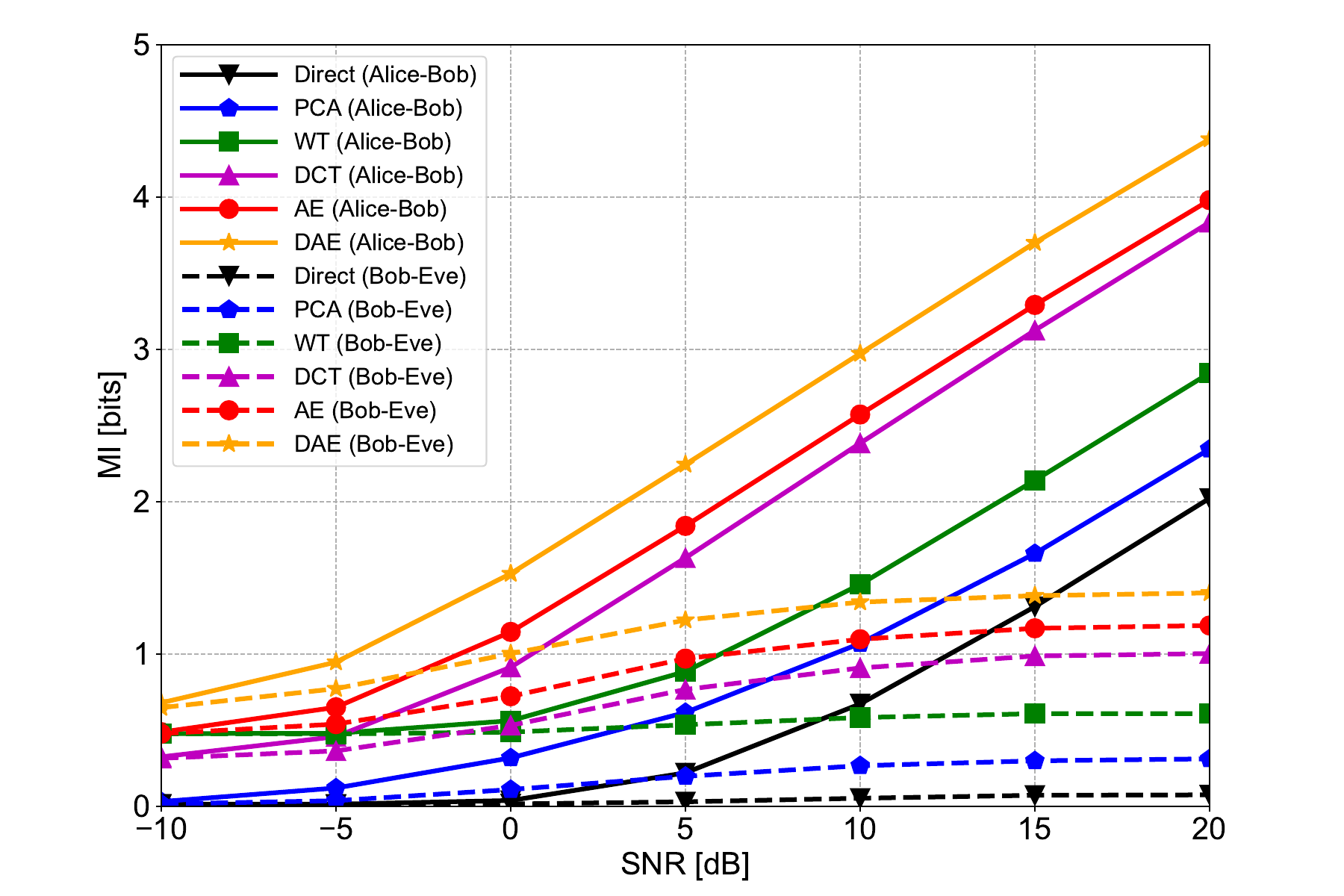}
\caption{MI of the proposed scheme and the five benchmarks with $\rho_{AB} = 1$ and $\rho_{BE} = 0.6$.}
\label{fig:mi_fig1}
\end{figure}

Fig.~\ref{fig:mi_fig1} shows MI between Alice and Bob and that between Bob and Eve, where we have the correlations of $\rho_{AB} = 1$ and $\rho_{BE} = 0.6$. As for the MI between Alice and Bob, the proposed scheme outperforms the benchmark schemes in the entire SNR regime. However, it is also observed that the proposed scheme's MI between Bob and Eve is higher than those of the benchmarks. Hence, the resultant secrecy performance has to be evaluated in terms of SKC, which will be shown later in Section~\ref{sec:Evaluation of SKC}

\begin{figure}[!t]
\centering
\includegraphics[width=.8\columnwidth]{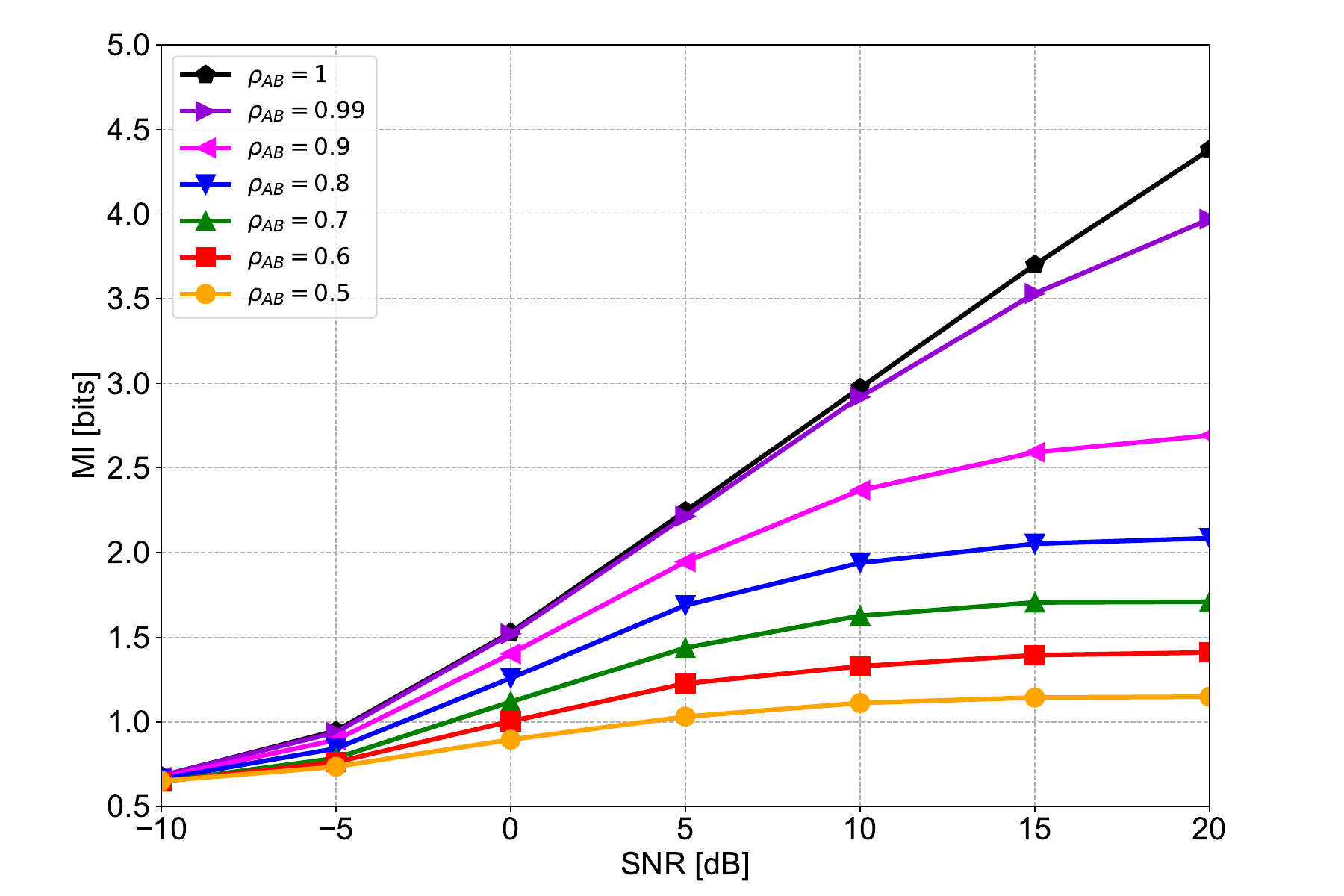}
\caption{MI of the proposed scheme, where the channel correlation between Alice and Bob is varied from $\rho_{AB}=0.5$ to $1$.}
\label{fig:mi_fig2}
\end{figure}

Fig.~\ref{fig:mi_fig2} illustrates the proposed scheme's MI between Alice and Bob, where the channel correlation coefficient is varied from $\rho_{AB}=0.5$ to $1$.
For $\rho_{AB} = 1$ or $0.99$, the MI increases with the increase of the SNR. However, upon decreasing the correlation value as $\rho_{AB} \le 0.7$, the MI exhibits clear saturation.
This may be due to the fact that the effects of channel mismatch are dominant.

\subsection{SKC Performance}
\label{sec:Evaluation of SKC}

\begin{figure}[!t]
\centering
\includegraphics[width=.8\columnwidth]{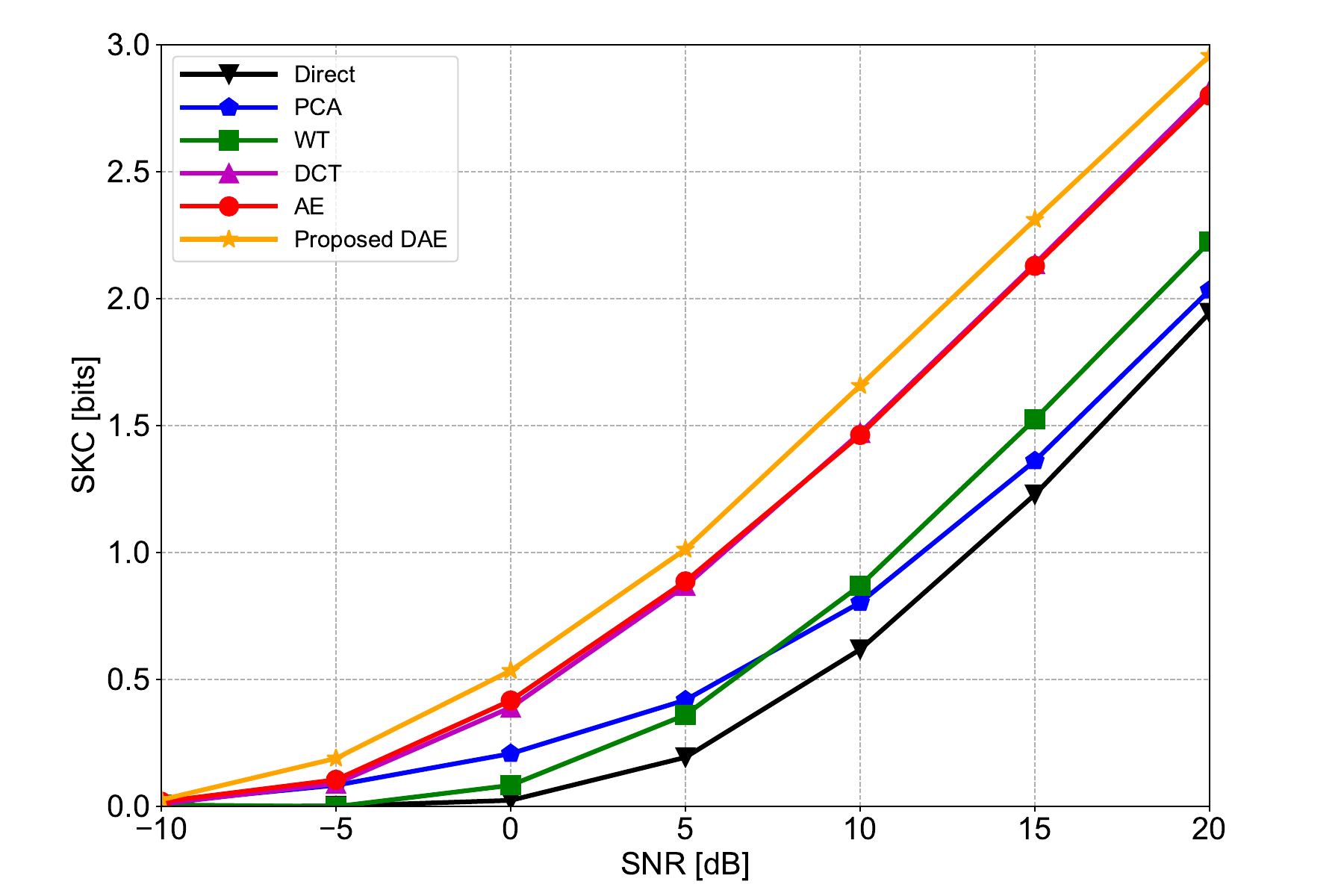}
\caption{SKC of the proposed scheme and the five benchmarks for $\rho_{AB} = 1$ and $\rho_{BE} = 0.6$.}
\label{fig:skc_fig1}
\end{figure}

To evaluate the achievable secrecy rate of the proposed scheme, which takes into account the information leakage to the eavesdropper, we provide the associated SKC performance~\cite{1055892}. 
Fig.~\ref{fig:skc_fig1} shows the SKC of the proposed scheme and the five benchmarks for $\rho_{AB} = 1$ and $\rho_{BE} = 0.6$. Over the entire SNRs, the proposed scheme outperforms the benchmarks. This confirms that the DAE is able to increase the MI between the legitimate users while limiting the information leakage to the eavesdropper.

\subsection{Randomness Test}
\label{sec:RandomnessTest}

To demonstrate the robustness of the cryptographic keys generated by the proposed scheme, we provide the rigorous statistical evaluations and the NIST SP 800-22 test suite~\cite{nist80022}. Here, 1,000 independent 1,536-bit keys are generated at the SNR $= 20$~dB. For each test, a $p$-value is calculated under the null hypothesis that the sequence is random. According to the standard NIST, a sequence has to attain the $p$-value higher than $\alpha = 0.01$.
\begin{table}[t]
\footnotesize
\caption{NIST Test Results}
\label{tab:nist}
\begin{center}
\renewcommand{\arraystretch}{1.2}
\begin{tabular}{lc} \hline
\textbf{Test Name} & \textbf{p-value} \\ \hline
Monobit frequency test & 1.0 \\ 
Block frequency test & 0.0574 \\ 
Runs test & 0.6314 \\ 
Longest run of ones test & 0.0959 \\ 
Discrete Fourier transform test & 0.3334 \\ 
Approximate entropy test & 0.0244 \\ 
Cumulative sums test & 0.2343 \\ \hline
\end{tabular}
\end{center}
\end{table}
More specifically, we consider seven tests, i.e., the monobit frequency test, the block frequency test, the runs test, the longest run of ones test, the discrete Fourier transform test, the approximate entropy test, and the cumulative sums test.

Table~\ref{tab:nist} shows the resultant average p-values for the seven tests, each exhibiting sufficiently higher than the significance level of $0.01$. This indicates that the sequences generated by the proposed DAE-aided SKG scheme do not possess any detectable statistical biases and successfully pass the suite of randomness tests.

\section{Conclusions}
\label{sec:Conclusions}
In this paper, we proposed the novel ML-based SKG framework, exploiting the offline DAE training and online SKG.
More specifically, the noise-contaminated estimated channels are recovered to construct a bit sequence reliably shared between the legitimate users, hence reducing the KDR and increasing the SKC. Our performance results demonstrated that the proposed SKG scheme outperforms the state-of-the-art benchmarks, such as the DWT, the DCT, the PCA, and the conventional AE schemes. Furthermore, the statistical tests with the NIST SP 800-22 suite confirmed that the generated key successfully passes several randomness criteria, validating the secure random key source.

\section*{Acknowledgement}
This work was supported in part by the National Institute of Information and Communications Technology (NICT), Japan, under Grant JPJ12368C00801, the Japan Science and Technology Agency (JST) FOREST (Grant JPMJFR2127), in part by the JST ASPIRE (Grant JPMJAP2345), and in part by the Japan Society for the Promotion of Science (JSPS) KAKENHI (Grants 23H00470, 23K22752, 24K21615).

\bibliographystyle{IEEEtran}
\bibliography{IEEEabrv,bib}

\end{document}